\documentclass[12pt,aip,superscriptaddress,preprint]{revtex4-1}

\usepackage{amsmath, txfonts} 
\usepackage{graphicx}
\usepackage{epstopdf}
\usepackage{verbatim}
			
\begin{document}

\newcommand{\subeff}{ \mathrm{eff} }
\newcommand{\subin}{ \mathrm{in} }
\newcommand{\subex}{ \mathrm{ex} }
\newcommand{\subsq}{ \mathrm{sq} }
\newcommand{\subtri}{ \mathrm{tri} }
\newcommand{\subT}{ \mathrm{T} }

\allowdisplaybreaks

\title{Acoustic metamaterial absorbers based on confined sonic crystals}
\author{Matthew D. Guild}
\email{mdguild@utexas.edu}
\affiliation{Grupo de Fen\'{o}menos Ondulatorios, Departamento de Ingenieria Electr\'{o}nica,
Universitat Polit\`{e}cnica de Val\`encia, Camino de vera s/n, E-46022 Valencia, Spain}
\author{Victor M. Garcia-Chocano}
\affiliation{Grupo de Fen\'{o}menos Ondulatorios, Departamento de Ingenieria Electr\'{o}nica,
Universitat Polit\`{e}cnica de Val\`encia, Camino de vera s/n, E-46022 Valencia, Spain}
\author{Weiwei Kan}
\affiliation{Department of Physics, Key Laboratory of Modern Acoustics, MOE, Institute of Acoustics, 
Nanjing University, Nanjing 210093, People's Republic of China}
\author{Jos\'{e} S\'{a}nchez-Dehesa}
\affiliation{Grupo de Fen\'{o}menos Ondulatorios, Departamento de Ingenieria Electr\'{o}nica,
Universitat Polit\`{e}cnica de Val\`encia, Camino de vera s/n, E-46022 Valencia, Spain}

\date{\today}

\begin{abstract}
Theoretical, numerical and experimental results examining thermoviscous losses in sonic crystals are presented in this work, enabling the fabrication and characterization of an acoustic metamaterial absorber with complex-valued anisotropic inertia.  The formulations developed can be written with no unknown or empirical coefficients, due to the structured lattice of the sonic crystals and organized layering scheme, and it is shown that higher filling fraction arrangements can be used to provide a large enhancement in the loss factor.  To accurately describe these structures in a realizable experimental configuration, confining structures are needed which modify the effective properties, due to the thermal and viscous boundary layer effects within the sonic crystal lattice.  Theoretical formulations are presented which describe the effects of these confined sonic crystals, both individually and as part of an acoustic metamaterial structure, and is demonstrated experimentally in an acoustic impedance tube.  It is observed that confined sonic crystals demonstrate an increase in the viscous losses and a reduction in the effective bulk modulus, enabling better acoustic absorber performance through improved impedance matching and enhanced absorption.
\end{abstract}

\maketitle

\section{Introduction} \label{Sec:Intro}

Sonic crystals, defined as periodic distributions of sound scatterers in a fluid or air background, have been proposed as structures for attenuating and filtering sound waves because of their acoustic bandgaps\cite{Dowling1992,Sigalas1992,Sanchezperez1998}. 
Their refractive properties, which were studied in the pioneering work of Kock and Harvey\cite{Kock1949} back in 1949, were later revisited and expanded by Cervera and coworkers\cite{Cervera2002}. These authors developed an acoustic lens for airborne sound by using a cluster of rigid rods with external lenticular shape. The lensing behavior was understood to result from the effective properties of the cluster that, at low frequencies, behaves like a homogeneous fluid with some given effective mass density and bulk modulus. In fact, it has been demonstrated that sonic crystals, with hexagonal and square symmetries, behaves like isotropic fluids whose effective parameters simply depend on the lattice filling fraction\cite{Torrent2006}.

Research on sonic crystals below the homogenization limit has been boosted in recent years due to the possibility of using them as artificial structures with extreme homogenized properties, referred to as acoustic metamaterials, behaving as broadband anisotropic fluids, or metafluids \cite{Torrent2008a}.  Moreover, acoustic metamaterials or metafluids with mass anisotropy are receiving increasing attention due to the extraordinary acoustic devices predicted from transformation acoustics, like acoustic cloaks and acoustic hyperlenses, which require anisotropic fluids as the principal ingredient\cite{Cummer2008,Torrent2008,Li2009}.  Several designs and a few experimental demonstrations of acoustic metamaterials with dynamical mass anisotropy have been reported in the last few years\cite{Torrent2008,Pendry2008,Popa2009,Torrent2010,Gumen2011,Zigoneanu2011}, which make use of a nonresonant microstructure to create the desired anisotropy.  
 
In most applications, acoustic metamaterials have been envisioned using ideal materials, with the presence of losses seen as a hinderance to the design. However, acoustic waves in fluids such as air or water have inherent losses which arise from thermal and viscous effects, and can be particularly pronounced for small structures such as those encountered in metamaterial applications.  Furthermore, for sound absorber applications, these losses can be significant, and are in fact necessary to achieving the goal of absorbing the acoustic energy.  A recent study analyzed the homogenized properties of periodically distributed elastic cylinders embedded in a viscous fluid\cite{ReyesAyona2012}, however the analysis was constrained by the condition of low filling fractions, where the sound absorbing effects are not significant unless the frequencies are very high, or the structures are very small.

Recently, there has been interest in using the losses within an acoustic metamaterial to provide an enhancement in the absorption, using resonant structures such as membranes and mass-spring-damper systems \cite{Naify2010, Yang2010, Hussein2013}.  However, such resonant absorption mechanisms are inherently narrowband, and thus there is a need for nonresonant high loss structures in achieving broadband acoustic metamaterial absorbers.  Sonic crystals consisting of rigid rods arranged in  a hexagonal lattice with a large filling fraction have been recently employed to dissipate broadband acoustic energy at the core of an omni-directional sound absorber, also known as an \emph{acoustic black hole}\cite{Climente2012}, though the authors did not examine the physical mechanisms of the observed lossy behavior.   

In this work, the use of lossy sonic crystals with high filling fractions will be examined to demonstrate its applicability for sound absorbers, and how acoustic metamaterials with complex-valued effective material properties can be created and implemented, allowing for anisotropy in both the sound absorption characteristics and the effective properties.  To accurately understand the behavior of these structures in a realizable experimental configuration, confining structures are needed which modify the effective properties, due to the thermal and viscous boundary layer effects within the sonic crystal lattice.  These confined sonic crystal arrangements are found to exhibit an increase in the losses due to the increase in the effective viscosity and a decrease in the bulk modulus due to a change from adiabatic to nearly isothermal conditions within the homogenized sonic crystal.  The behavior of these confined sonic crystal structures is formulated theoretically, and is demonstrated experimentally in an acoustic impedance tube.  Although the confining structure experimentally examined is due to the testing apparatus, the theoretical formulation is more general, and demonstrates the use of confined sonic crystals to facilitate the design and realization of soft acoustic metamaterials \cite{Brunet2013}, enabling better acoustic absorbers through improved impedance matching and enhanced acoustic and absorption properties.

The work performed here is described as follows. In Section~\ref{Sec:TheoryAnalysisSC2D}, the theoretical formulations and parametric characterization of two-dimensional (2D) sonic crystals with thermovisous losses are presented and verified with numerical simulations. The properties of these lossy sonic crystals in a complex-valued anisotropic acoustic metafluid are then formulated in Section~\ref{Sec:AnisoInertia2D}.  Theoretical formulations for confined sonic crystals are developed in Section~\ref{Sec:ConfinedTheorySC} and the experimental results are described in Section~\ref{Sec:Experiment}, followed by a summary of the findings in Section~\ref{Sec:Conclusion}.

\section{Two-dimensional sonic crystals} \label{Sec:TheoryAnalysisSC2D}

For thermoviscous fluids, the properties of the sonic crystal are dependent on the size of the thermal and viscous boundary layers relative to that of the cylinder and lattice dimensions.  In particular, an expression for the effective homogenized properties is sought for a lattice of cylinders which are non-interacting, both fluid dynamically (i.e. boundary layers which do not touch) and acoustically (neglecting multiple scattering effects).  Extensive work has been performed on the topic of porous media, and detailed models have been developed to describe such systems.  The specific formulations in each case depend on the configuration of the microstructure.  Two-dimensional sonic crystals, which consist of parallel cylinders in a structured lattice, represent an idealized arrangement of a fibrous porous media, and previous work on such fibrous porous materials can provide a basis for development of a model for lossy sonic crystals.  A theoretical formulation for 2D lossy sonic crystals is presented in Section~\ref{Sec:TheorySC2D}, from which a nondimensional parameter space is developed and discussed in Section~\ref{Sec:ParamSpace2D}.  The theoretical results are then compared and verified with Comsol multiphysics simulations in Section~\ref{Sec:ComsolSC2D}.

\subsection{Theoretical formulation for 2D sonic crystals} \label{Sec:TheorySC2D}
\begin{figure}[t!]
	\includegraphics[width=0.99\columnwidth, height=0.7\textheight, keepaspectratio]{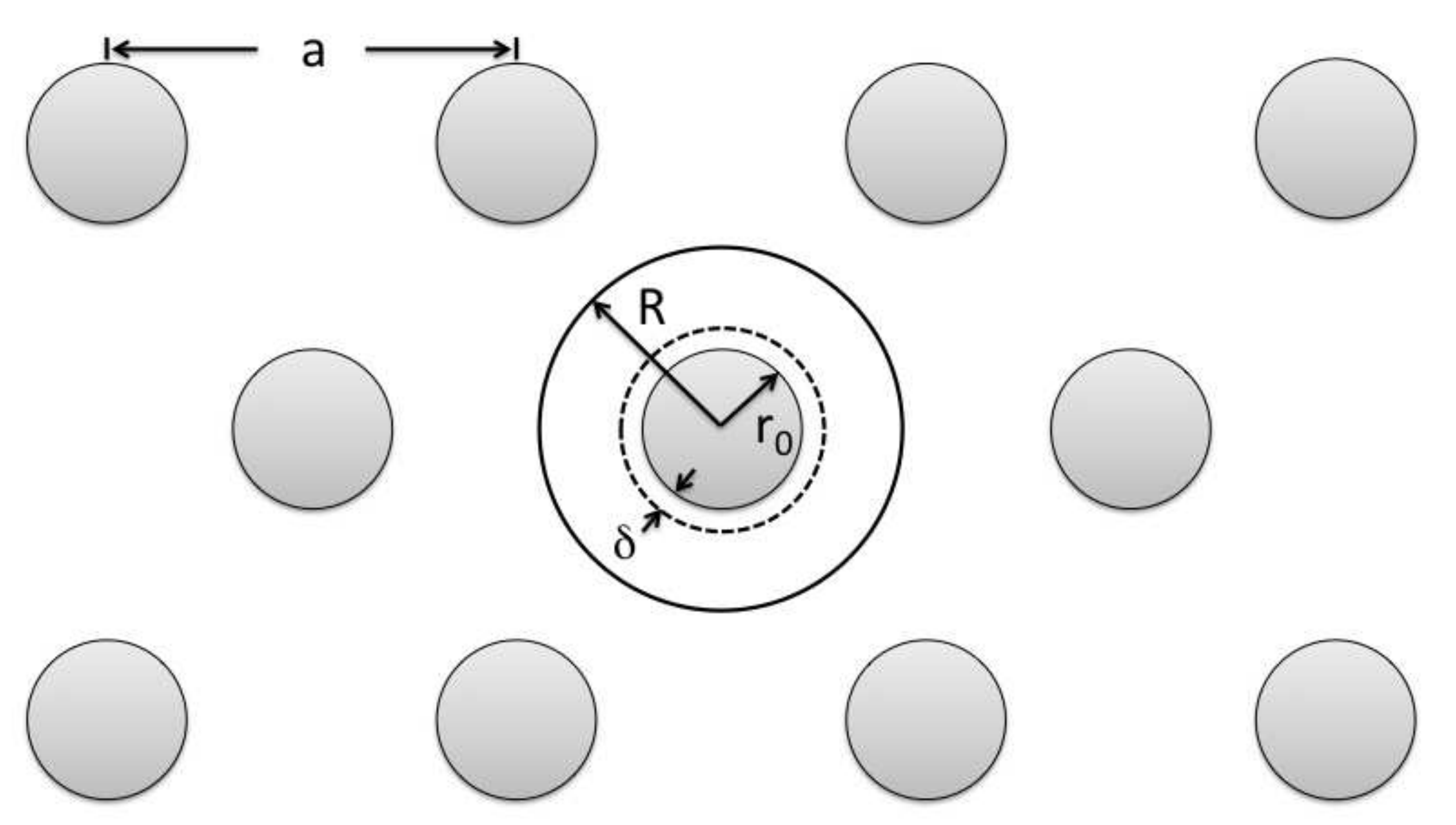}
	\caption{Geometry for a sonic crystal with lattice parameter $a$, cylinder radius $r_{0}$, and representative cell of radius $R$.}
	\label{Fig:LatticeGeom}
\end{figure}

The general form of the bulk density for rigid fibrous media consisting of parallel cylinders, such as the configuration illustrated in FIG.~\ref{Fig:LatticeGeom}, can be expressed as \cite{Allard1992}
\begin{equation} \label{Eq:Rho_FibrousMedia}
	\rho_{\mathrm{eff}} = \rho_{0} \frac{\bar{\alpha}}{1 \!-\! f},
\end{equation}
\noindent where $\rho_{0}$ is the density of the host fluid, $f$ is the filling fraction, and $\bar{\alpha}$ is dynamic tortuosity given by \cite{Johnson1987}
\begin{equation} \label{Eq:Alpha_FibrousMedia}
	\bar{\alpha} = \alpha_{\infty} \left[ 1 \!+\! \frac{\bar{F}}{j \bar{\omega}} \right],
\end{equation}
\noindent with $\alpha_{\infty}$ denoting the high frequency limit of the tortuosity and the functions $\bar{F}$ and $\bar{\omega}$ defined as
\begin{align}
	\bar{F} &= \sqrt{ 1 \!+\! j \frac{1}{2}\bar{\omega} M }, \label{Eq:Fbar_FibrousMedia} \displaybreak[0] \\
	\bar{\omega} &= \frac{\omega \rho_{0} \alpha_{\infty}}{(1 \!-\! f) \sigma}, \label{Eq:Omega_FibrousMedia} \displaybreak[0] \\
	M &= \frac{8 \alpha_{\infty} \eta}{(1 \!-\! f) \sigma \Lambda^{2}}. \label{Eq:M_FibrousMedia}
\end{align}
From these equations, it can be seen that the effects of the losses arise from the dynamic viscosity $\eta$, the static flow resistivity $\sigma$ and the characteristic viscous length $\Lambda$, which is a viscous parameter defined by Johnson \emph{et al.}\cite{Johnson1987}.  For the density of a porous medium, the losses arise from viscous effects, and result in an effective density which contains both a real and imaginary part.

For a lattice of rigid parallel cylinders embedded in an ideal gas, like that illustrated in FIG.~\ref{Fig:LatticeGeom}, the bulk modulus can be written as
\begin{equation} \label{Eq:BulkMod_FibrousMedia}
	\kappa_{\mathrm{eff}} = \frac{\gamma P_{0}}{(1 \!-\! f)} \frac{1}{C_{\mathrm{fiber}}},
\end{equation}
\noindent where $\gamma$ is the ratio of specific heats, $P_{0}$ is the ambient static pressure, and $f$ is the filling fraction.  The sound speed can be determined from Equations~(\ref{Eq:Rho_FibrousMedia}) and (\ref{Eq:BulkMod_FibrousMedia}) by $c_{\mathrm{eff}} \!=\! \sqrt{\kappa_{\mathrm{eff}}/\rho_{\mathrm{eff}}}$.  In Equation~(\ref{Eq:BulkMod_FibrousMedia}), $C_{\mathrm{fiber}}$ is the dynamic compressibility, which can be determined based on the thermal boundary conditions.

For thermally conducting fibers, the boundary condition on the temperature change, $T$, at the fiber edge, $r \!=\! r_{0}$, with $e^{j\omega t}$ time dependence is given by \cite{Tarnow1996}
\begin{equation} \label{Eq:ThermCondBC}
	2 \pi r_{0} \tau \! \left. \frac{\partial T}{\partial r} \right|_{r \!=\! r_{0}} \!\!  = j \omega \, \pi r_{0}^{2} \, \rho_{\mathrm{fiber}} \mathrm{C}_{\mathrm{p,fiber}} T(r_{0}),
\end{equation} 
\noindent where $\tau$ is the thermal conductivity of the fluid (air), $\rho_{\mathrm{fiber}}$ is the mass density of the fiber, and $\mathrm{C}_{\mathrm{p,fiber}}$ is the specific heat capacity of the fiber.  For thermally conducting fibers in air, the density of typical solid materials is several orders of magnitude larger than that of air, and therefore the term on the right hand side of Equation~(\ref{Eq:ThermCondBC}) will dominate, giving nearly isothermal conditions, $T(r_{0}) \approx 0$.  Similarly, when the spacing  between the cylinders is large compared with the thermal boundary layer (corresponding to either relatively high frequencies or low filling fractions), any thermal interaction with the surrounding cylinders can be neglected.  In this case, the thermal boundary condition at the outer radius $R$ is adiabatic, 
\begin{equation} \label{Eq:AdiabaticBC}
	\left. \frac{\partial T}{\partial r} \right|_{r \!=\! R} \!\!  = 0.
\end{equation}

Applying these thermal boundary conditions, the dynamic compressibility can be obtained, \cite{Tarnow1996}
\begin{align}
	C_{\mathrm{fiber}} &= 1 - (\gamma-1) \frac{2 f}{(1 \!-\! f)} \bar{H} ,\label{Eq:C_FibrousMedia} \displaybreak[0] \\
	\bar{H} &= \frac{1}{k_{T}r_{0}} \frac{ \left[ J_{1}(k_{T}r_{0}) H_{1}^{(2)}(k_{T}R) \!-\! J_{1}(k_{T}R) H_{1}^{(2)}(k_{T}r_{0}) \right]  }{  \left[ J_{0}(k_{T}r_{0}) H_{1}^{(2)}(k_{T}R) \!-\! J_{1}(k_{T}R) H_{0}^{(2)}(k_{T}r_{0}) \right] }, \label{Eq:H_bar2D}
\end{align}
\noindent where $H_{m}^{(2)}$ is the $m^{\mathrm{th}}$ order Hankel function of the second kind, $r_{0}$ is the cylinder radius, $R$ is the radius defined by the filling fraction $f = (r_{0}/R)^{2}$, and the thermal wavenumber $k_{\subT}$ is
\begin{equation} \label{Eq:kT_FibrousMedia}
	k_{\subT} = (1 \!-\! j) \sqrt{ \frac{\omega \rho_{0} \mathrm{Pr}}{2 \eta} } = (1 \!-\! j)  \frac{\sqrt{\mathrm{Pr}}}{\delta} = (1 \!-\! j)  \frac{1}{\delta'}.
\end{equation}
Note that the thermal wavenumber can be written in terms of the viscous boundary layer thickness $\delta$ or the thermal boundary layer thickness $\delta'$,
\begin{equation} \label{Eq:delta_therm}
	\delta' = \sqrt{\frac{2 \eta}{\omega \rho_{0} \mathrm{Pr}} } = \frac{\delta}{\sqrt{\mathrm{Pr}}},
\end{equation}
\noindent and therefore, the reduced thermal frequency $k_{\subT}r_{0}$ can be expressed as
\begin{equation} \label{Eq:kr_SC}
	k_{\subT}r_{0} = (1 \!-\! j)  \sqrt{\mathrm{Pr}}\left( \frac{\delta}{r_{0}} \right)^{-1}.
\end{equation}
It can be seen that the effects of the losses arise from both viscous and thermal effects, leading to a complex value for the bulk modulus.  The thermal effects are quantified by the the Prandtl number $\mathrm{Pr}$, a dimensionless parameter which relates the contributes of the thermal relative to the viscous effects,
\begin{equation} \label{Eq:PrandtlDef}
	\mathrm{Pr} = \frac{ \eta \mathrm{C}_{\mathrm{p}} }{\tau},
\end{equation}
\noindent where $\eta$ is the viscosity, $\mathrm{C}_{\mathrm{p}}$ is the specific heat capacity and $\tau$ is the thermal conductivity.

\subsubsection{Formulation of relevant model parameters}
Due to the random nature of the fibrous media which has traditionally been examined, the existing literature has focused on the case of low volume fractions (often on the order of a few percent), using parameters which often require experimental characterization of specific samples since the precise microstructure is not known.  For such naturally occurring materials, higher volume fractions without a precisely arranged microstructure will tend to clump and intersect, creating what would essentially appear like pores.  As a result, the situations of moderate to high concentrations of fibrous porous media have typically been neglected.  For lossy sonic crystals, however, these closely packed arrangements are of particular interest, and represent the exact microstructure that one wishes to examine.  The three relevant model parameters characterizing the viscous and thermal effects of the lattice structure are: the viscous characteristic length $\Lambda$, the tortuosity $\alpha_{\infty}$ and the flow resistivity $\sigma$.  An additional parameter, the thermal characteristic length $\Lambda'$, will also be presented, which will be shown to serve as an appropriate length scale for quantifying the thermal effects.

The viscous characteristic length is a metric of the viscous effects proposed by Johnson \emph{et al.} \cite{Johnson1987}.  Evaluation of this quantity analytically for the viscous fluid flow around a rigid cylinder yields \cite{Allard1992}
\begin{equation} \label{Eq:LambdaDef}
	\Lambda = \frac{r_{0}}{2 f}(1 - f^{2}).
\end{equation}
Note that for small filling fractions, Equation~(\ref{Eq:LambdaDef}) yields $\Lambda \! \approx \! r_{0} / (2 f)$, the same as that obtained by Allard and Champoux  \cite{Allard1992}.  However, for the moderate to high filling fractions that can be achieved using sonic crystals, the higher precision of the exact expression given by Equation~(\ref{Eq:LambdaDef}) is necessary to accurately describe the acoustic performance.

In a similar form to that of the viscous characteristic length, the thermal effects can be quantified by the characteristic thermal length $\Lambda'$, in addition to the Prandtl number \cite{Champoux1991}.  Following a similar process as above, an expression for $\Lambda'$ can be obtained \cite{VenegasThesis2011}
\begin{equation} \label{Eq:LambdaPrimeDef}
	\Lambda' = \frac{r_{0}}{f}(1 - f).
\end{equation}
As in the case of the viscous characteristic length, retaining the higher order terms with respect to the filling fraction is necessary when considering more compact configurations made possible by the use of sonic crystal lattices.

In the context of sonic crystals, the unique homogenized bulk properties arise from dynamic effects, and thus for the static flow resistivity it is more appropriate to consider this as a quasi-static condition of low but non-zero oscillatory flow.  For a lattice of parallel rigid cylinders, an expression for the flow resistivity of a structured lattice has been derived by Tarnow \cite{Tarnow1996a}
\begin{equation} \label{Eq:SigmaDef}
	\sigma = \frac{4 f \eta}{r_{0}^{2} \left[ -\frac{1}{2}\ln f -\frac{3}{4} + f - \frac{1}{4} f^{2}\right]},
\end{equation}
\noindent which is equivalent to the earlier solution derived for a square lattice following a similar approach by Kuwabara\cite{Kuwabara1959}.  In both cases, the solution was developed by using a circular representative volume of fluid surrounding each cylinder (illustrated in FIG.~\ref{Fig:LatticeGeom}), and assuming free conditions at the boundary each cell.

For parallel cylindrical lattices, Tournat \emph{et al.}\cite{Tournat2004} derived an expression for the tortuosity, $\alpha_{\infty} = 1 \!+\! f$.  Although originally developed as an approximate solution valid only for small filling fractions, this solution holds for all filling fractions in the absence of multiple scattering effects.  This can be seen by comparing Equation~(\ref{Eq:Rho_FibrousMedia}) in the limit of zero viscosity to the lossless quasi-static dynamic density for a sonic crystal, given by \cite{Torrent2006,Martin2010}
\begin{equation} \label{Eq:RhoEff_NoLosses}
	\rho_{\mathrm{eff,0}} = \rho_{0} \frac{1 \!+\! f}{1 \!-\! f}.
\end{equation}

\subsubsection{Effective density of a 2D sonic crystal with losses}
With the expressions for $\Lambda$, $\sigma$ and $\alpha_{\infty}$ presented above, the complex effective density given by Equation~(\ref{Eq:Rho_FibrousMedia}) for a sonic crystal with viscous losses can be written as
\begin{align}
	\rho_{\mathrm{eff}} &= \rho_{0} \left( \frac{1 \!+\! f}{1 \!-\! f} \right)\left[ 1 - j \frac{\bar{F}_{\mathrm{sc}}}{\bar{\omega}_{\mathrm{sc}}}\right], \label{Eq:Rho_SC} \displaybreak[0] \\
	\bar{F}_{\mathrm{sc}} &= \sqrt{ 1 \!+\! j \frac{1}{2}\bar{\omega}_{\mathrm{sc}} M_{\mathrm{sc}} }, \label{Eq:Fbar_SC} \displaybreak[0] \\
	\bar{\omega}_{\mathrm{sc}} &= \frac{1}{2 f (\frac{\delta}{r_{0}})^{2}} \left( \frac{1 \!+\! f}{1 \!-\! f} \right) \left[ -\frac{1}{2} \ln f - \frac{3}{4} + f -\frac{1}{4}f^{2} \right], \label{Eq:Omega_SC} \displaybreak[0] \\
	M_{\mathrm{sc}} &= \frac{8 f}{\left(1 \!-\! f^{2}\right)^{2}} \! \left( \frac{1 \!+\! f}{1 \!-\! f} \right) \! \left[ -\frac{1}{2} \ln f - \frac{3}{4} + f -\frac{1}{4}f^{2} \right]. \label{Eq:M_SC}
\end{align}
From these equations, it is clear that besides the host fluid density, the only parameters that affect the density are the filling fraction $f$ and the ratio of the viscous boundary layer thickness to the cylinder radius $\delta/r_{0}$.  From the definition of $\delta$ given in Equation~(\ref{Eq:delta_therm}), it can be seen that this term includes all the relevant viscous effects and the frequency dependence.  Unlike unstructured porous media, which require estimated or experimentally determined scaling parameters \cite{Attenborough1982}, there are no free parameters required for modeling the bulk effective properties of a lossy sonic crystal.  Therefore, the expression presented above for the complex density of a sonic crystal is an explicit expression in terms of the host density and filling fraction, with all the viscous and dispersive effects accounted for by a single dimensionless parameter, $\delta/r_{0}$, which can be calculated based on the frequency and the properties of the viscous host fluid.

In the limiting case where the viscous boundary layer is thin, $(\delta/r_{0}) \! \ll \! 1$, the expressions for the complex density of a lossy sonic crystal can be simplified.  For $(\delta/r_{0}) \! \ll \! 1$, this implies that $\bar{\omega}_{\mathrm{sc}} \! \gg \! 1$, and therefore the expression for the complex density becomes
\begin{equation} \label{Eq:Rho_SC_HighFreq}
	\rho_{\mathrm{eff}} \approx \rho_{0} \left( \frac{1 \!+\! f}{1 \!-\! f} \right)\left[ 1 \!+\! \frac{\delta}{\Lambda} \! \left( 1 \!-\! j \right) \right], \quad \frac{\delta}{r_{0}} \!\ll\! 1,
\end{equation}
\noindent where use of Equation~(\ref{Eq:LambdaDef}) gives
\begin{equation} \label{Eq:Delta2Lambda}
	\frac{\delta}{\Lambda} =  \left( \frac{2f}{1 \!-\! f^{2}} \right) \frac{\delta}{r_{0}}.
\end{equation}

Based on Equation~(\ref{Eq:Rho_SC_HighFreq}), it is apparent that the presence of viscosity affects both the real and imaginary parts of the density.  The imaginary part, which is identically equal to zero for an inviscid fluid, is linearly proportional to $\delta/r_{0}$ for small values of $\delta/r_{0}$.  For the real part, there is a viscous term which is equal in magnitude to the imaginary part, in addition to the nominal value for the lossless case.  Since this additional term is always positive, this means that viscous effects will lead to an \emph{increase} in the real part of the density above that of the nominal (lossless) case.  Furthermore, this viscous term varies with the filling fraction as $2 f / (1 \!-\! f^{2})$, an expression that can be equal to unity or greater for moderate to large filling fractions, so the increase in the real part of the density (as well as the imaginary part) can be significant.

\subsubsection{Effective bulk modulus of a 2D sonic crystal with losses}
In a similar manner, Equations~(\ref{Eq:BulkMod_FibrousMedia})-(\ref{Eq:kr_SC}) can be used to express the bulk modulus of a sonic crystal with viscous losses,
\begin{align}
	\kappa_{\mathrm{eff}} &= \frac{\gamma P_{0}}{(1 \!-\! f)} \frac{1}{C_{\mathrm{sc}}}, \label{Eq:BulkMod_SC} \displaybreak[0] \\
	C_{\mathrm{sc}} &= 1 - (\gamma-1) \frac{2 f}{(1 \!-\! f)} \bar{H}. \label{Eq:C_SC} 
\end{align}
As with the complex density, the complex bulk modulus is a function only of the properties of the host fluid (including thermal properties), the filling fraction and the parameter $\delta/r_{0}$.

To determine an expression for the bulk modulus when the viscous boundary layer is thin, Equations~(\ref{Eq:H_bar2D}) and (\ref{Eq:BulkMod_SC}) can be simplified by using the large argument (high frequency) limits of the Bessel functions, which yields an approximate solution for the bulk modulus, \cite{Tarnow1996}
\begin{equation} \label{Eq:BulkMod_SC_HighFreq}
	\kappa_{\mathrm{eff}} \approx \frac{\gamma P_{0}}{ (1 \!-\! f) \left[ 1 + (\gamma-1) \frac{\delta'}{\Lambda'} (1 \!-\! j) \right]}, \quad \frac{\delta}{r_{0}} \ll 1,
\end{equation}
\noindent where use of Equations~(\ref{Eq:delta_therm}) and (\ref{Eq:LambdaPrimeDef}) gives
\begin{equation} \label{Eq:Delta2LambdaPrime}
	\frac{\delta'}{\Lambda'} = \frac{1}{\sqrt{\mathrm{Pr}}} \left( \frac{f}{1 \!-\! f} \right) \frac{\delta}{r_{0}}.
\end{equation}
As with the density, it is apparent from Equations~(\ref{Eq:BulkMod_SC_HighFreq}) and (\ref{Eq:Delta2LambdaPrime}) that both the real and imaginary parts of the bulk modulus are affected by viscosity, containing only linear terms of $\delta/r_{0}$.  In contrast with the density, though, the viscous term appears in the denominator.  The term in brackets is always greater than unity for non-zero viscosity, leading to a \emph{decrease} in the real part of the bulk modulus compared to the nominal (lossless) case.

\subsection{Parametric representation of lossy sonic crystals} \label{Sec:ParamSpace2D}
From Equations~(\ref{Eq:H_bar2D}), (\ref{Eq:Rho_SC})--(\ref{Eq:M_SC}), (\ref{Eq:BulkMod_SC}) and (\ref{Eq:C_SC}), theoretical values for the complex density and bulk modulus can be obtained.  These expressions do not contain any empirically derived coefficients, and for a given host fluid can completely describe any combination of lattice geometries and frequency using two independent parameters: the filling fraction $f$ and the normalized viscous boundary layer thickness, $\delta/r_{0}$.  Therefore, it is possible to create parametric plots of the effective sonic crystal properties, which can encompass the entire range of possible effective properties for 2D sonic crystals with thermoviscous losses, for a given host fluid.  Use of such plots allow for the design and interpretation of sonic crystal effective properties when thermoviscous losses are present, and enable one to better characterize the potential absorption properties of a sonic crystal.

\begin{figure}[t!]
	\includegraphics[width=0.99\columnwidth, height=0.7\textheight, keepaspectratio]{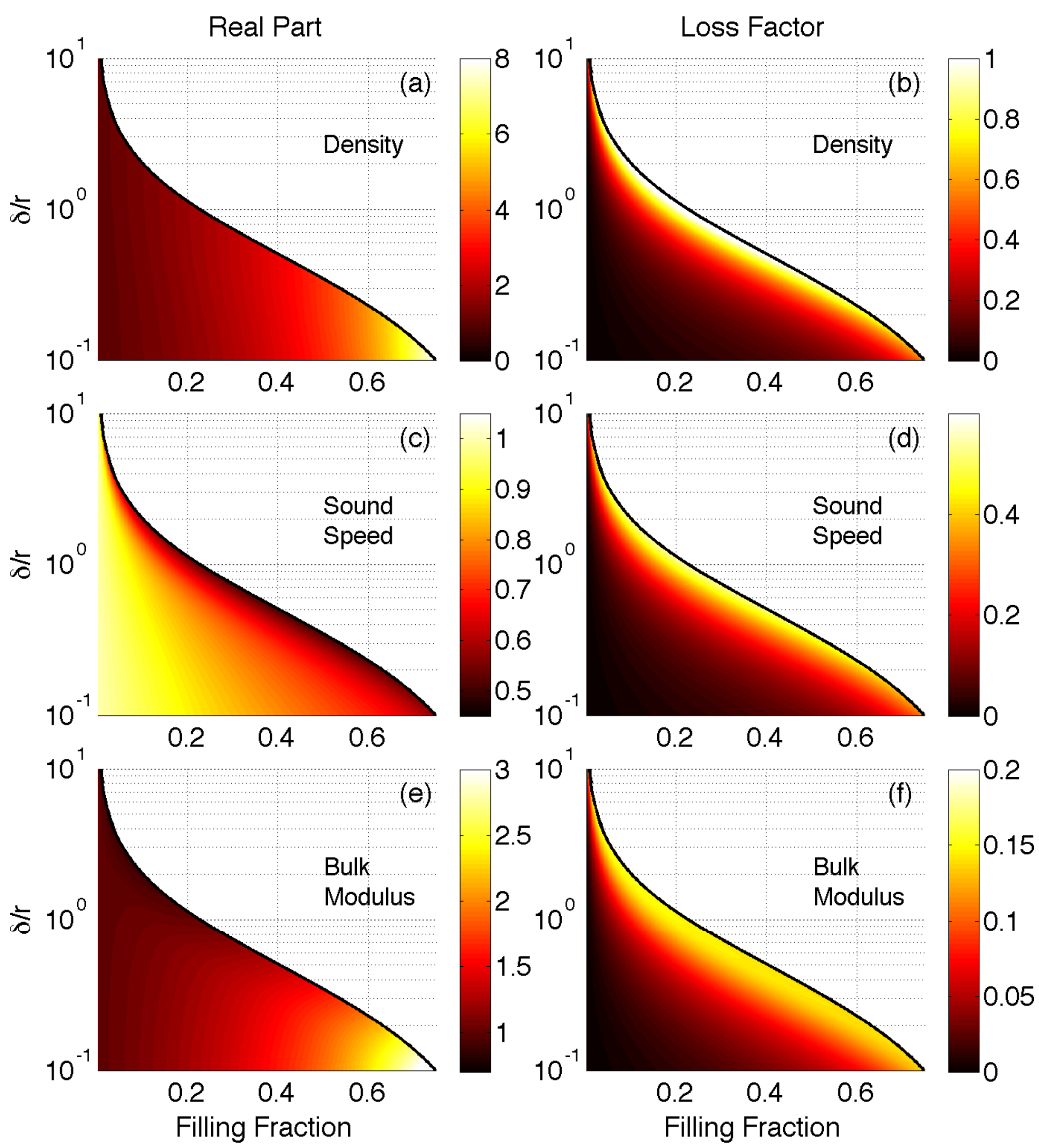}
	\caption{(Color online) Intensity plots of the real part and loss factor for the complex effective density, sound speed and bulk modulus.  The color scale of the plots ranges from low values (dark) to high values (light). }
	\label{Fig:SurfPlot}
\end{figure}

Parametric plots versus filling fraction and $\delta/r_{0}$ are illustrated in FIG.~\ref{Fig:SurfPlot} for the complex density, sound speed and bulk modulus of a 2D sonic crystal in air.  The parameter space has been limited to values where the thermal and viscous boundary layers are sufficiently small so as they do not touch the boundary layer of the adjacent cylinders.  The limiting case where the boundary layers touch is denoted by a solid black line.  FIG.~\ref{Fig:SurfPlot}(a), (c) and (e) shows the real part of the property and FIG.~\ref{Fig:SurfPlot}(b), (d) and (f) shows the loss factor (imaginary part divided by the real part) on a color scale, ranging from low values (dark) to high values (light).  Note that while fibrous porous materials have been extensively utilized for sound absorbing applications, these have traditionally been limited to low filling fractions, on the order of a few percent, which represents the left-most region of the plots.

Expanding the parameter space to include the higher filling fractions made possible by the structured lattice of the sonic crystals, one can identify several desirable features which could be utilized for acoustic absorbers.  In particular, it is observed that there is a broad region across the moderate to high filling fractions where the loss factor is large, and in the case of the density approaches unity, compared with very small values for the region covered by traditional fibrous porous absorbers.  In addition, from FIG.~\ref{Fig:SurfPlot}(b) it can be seen that significant reductions in the real part of the sound speed, which represents the speed of the wave through the homogenized sonic crystal, occur at moderate to high filling fractions.  Although this does not change the absorption per cycle, it does affect the wavelength of the sound passing through the absorber.  Decreasing the sound speed, as shown in FIG.~\ref{Fig:SurfPlot}(b), will decrease the wavelength, and therefore lead to an absorber which appears acoustically ``thicker" and thereby increasing the total absorption.

\subsection{Comparison of results with Comsol} \label{Sec:ComsolSC2D}

\begin{figure}[t!]
	\includegraphics[width=0.99\columnwidth, height=0.7\textheight, keepaspectratio]{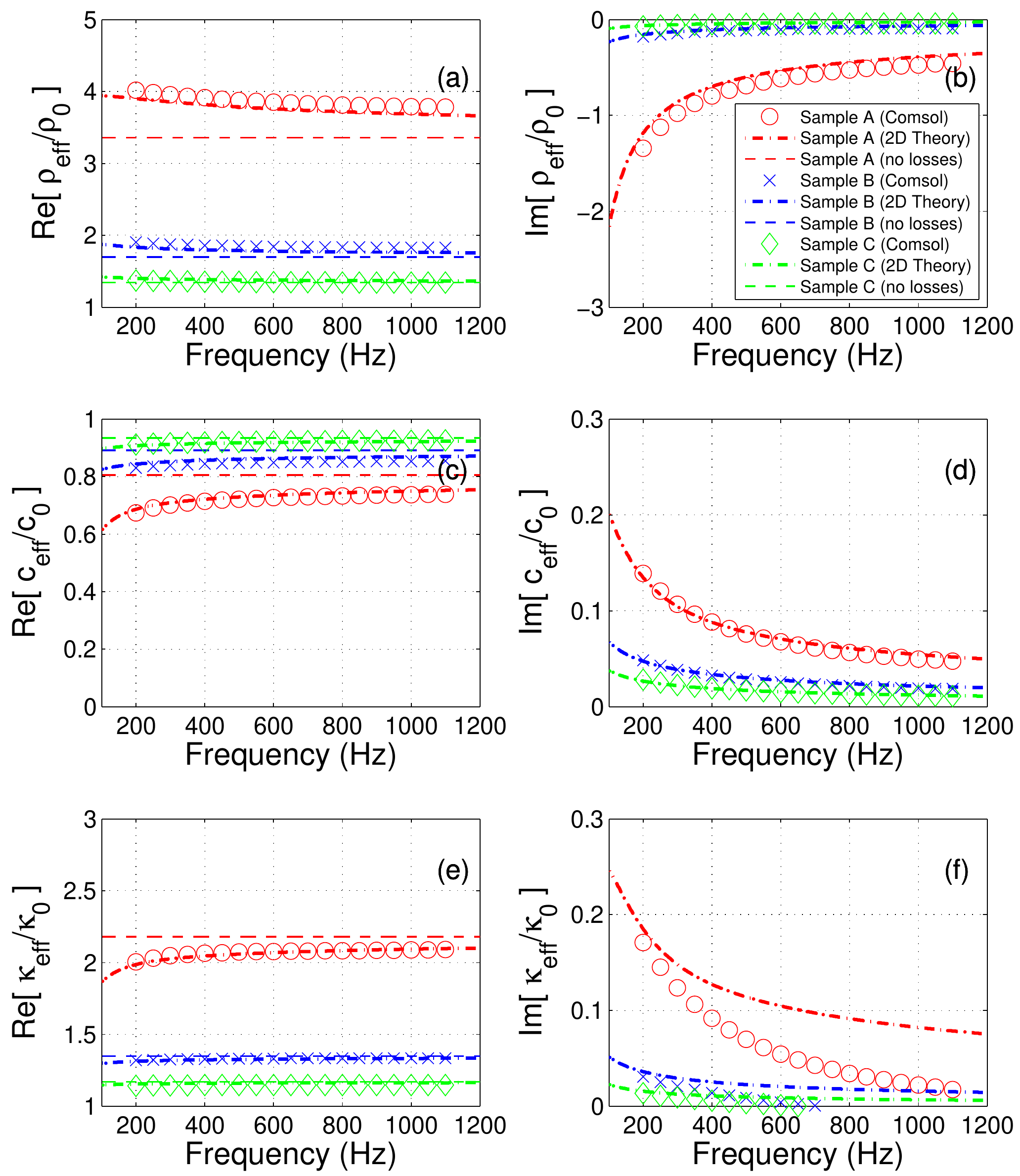}
	\caption{(Color online) Comparison of theoretical results and Comsol for the real and imaginary parts of the density and bulk modulus of a sonic crystal with $r_{0} \!=\! 1$ mm, for Sample A, B, and C.}
	\label{Fig:ModelComp2D}
\end{figure}

To verify the theoretical formulation developed in Section~\ref{Sec:TheorySC2D}, the complex density and bulk modulus are compared with Comsol simulations.  In the Comsol models, the cylinders are assumed to be rigid, and a thermoviscous host fluid with the properties of air is used.  Although the dimensionless parameter $\delta/r_{0}$ is utilized for the theoretical analysis, the use of Comsol requires specific dimensions and a corresponding frequency range for the acoustical modeling to be performed.  Based on the definition of $\delta$ given in Equation~(\ref{Eq:delta_therm}), $\delta/r_{0}$ can be calculated for a specific fluid (in this case air) and frequency range and the cylinder radius $r_{0}$.  

Results calculated from Equations~(\ref{Eq:Rho_SC}) and (\ref{Eq:BulkMod_SC}) are compared with Comsol simulations in FIG.~\ref{Fig:ModelComp2D}, for Samples A, B, and C, the dimensions of which are listed in Table~\ref{Tab:SampleABC}.  For reference, effective properties for the lossless case are shown in FIG.~\ref{Fig:ModelComp2D}(a), (c) and (e) and denoted by a dashed line.  In this figure, it is clear that there is excellent agreement between the theoretical model developed here and the Comsol data, for both the real and imaginary parts of the density, sound speed and bulk modulus.  Conversely, the effective properties for the lossless cases fail to capture the trends in the data for even the real part of the effective properties as a function of frequency, and the overall magnitude deviates from that of either the theory with losses or Comsol at higher filling fractions.   Based on these results, the theoretical formulation with losses provides a relatively simple yet accurate explicit formulation for the effective properties of a lossy sonic crystal structure.

\begin{table}[t!]
	\begin{ruledtabular}
		\begin{tabular}{ccccc}
			\bf{Sample} & $r_{0}$ (mm) & $a$ (mm) & Length, $L$ (mm) & Filling fraction, $f$ \\ \hline
			A & $1.0$ & $2.5$ & $42.5$ & $0.541$ \\
			B & $1.0$ & $3.8$ & $43.1$ & $0.234$ \\
			C & $1.0$ & $5.0$ & $45.0$ & $0.134$
		\end{tabular}
	\end{ruledtabular}
	\caption{Lattice properties for the three sonic crystal samples examined in this work.  The nominal sheath thickness, $l_{\mathrm{sheath}}$, and sheath lip, $l_{\mathrm{lip}}$, in each case was $0.5$ mm and $1$ mm, respectively.}
	\label{Tab:SampleABC}
\end{table}

\section{Acoustic metamaterial using alternating lossy sonic crystal layers} \label{Sec:AnisoInertia2D}

Recent work has examined anisotropic acoustic metamaterials theoretically, numerically and experimentally \cite{Gumen2011, Zigoneanu2011}.  Despite these thorough investigations and demonstrations of realizable structures, such works have neglected thermoviscous losses due to a primary focus on broadband, nonresonant acoustic metamaterials which operate without any appreciable losses.  Comparison of theoretical results with Comsol simulations for 2D sonic crystals (presented in FIG.~\ref{Fig:ModelComp2D}) shows that significant differences were observed between effective properties obtained assuming a lossless host fluid and those which include thermoviscous losses.  These differences resulted in a non-zero (and at some frequencies quite large) imaginary part, and also resulted in incorrect trends predicted by the lossless theory, including underestimating the real part of the density and overestimating the real part of the bulk modulus.  Therefore, in the following sections an analysis of an anisotropic acoustic metamaterial will be performed.  The theoretical framework for this will be discussed in Section~\ref{Sec:AnisoInertiaTheory}, in which the anisotropic acoustic metamaterial will be treated as a system of alternating effective fluid layers, with the properties of each effective fluid layer simply being the complex effective properties determined from the homogenization process of a uniform sonic crystal.  In Section~\ref{Sec:AnisoInertiaComsol}, the theoretical formulation is compared with 2D Comsol simulations with thermoviscous losses for a realizable configuration.

\subsection{Theoretical formulation} \label{Sec:AnisoInertiaTheory}

\begin{figure}[t!]
	\includegraphics[width=0.99\columnwidth, height=0.5\textheight, keepaspectratio]{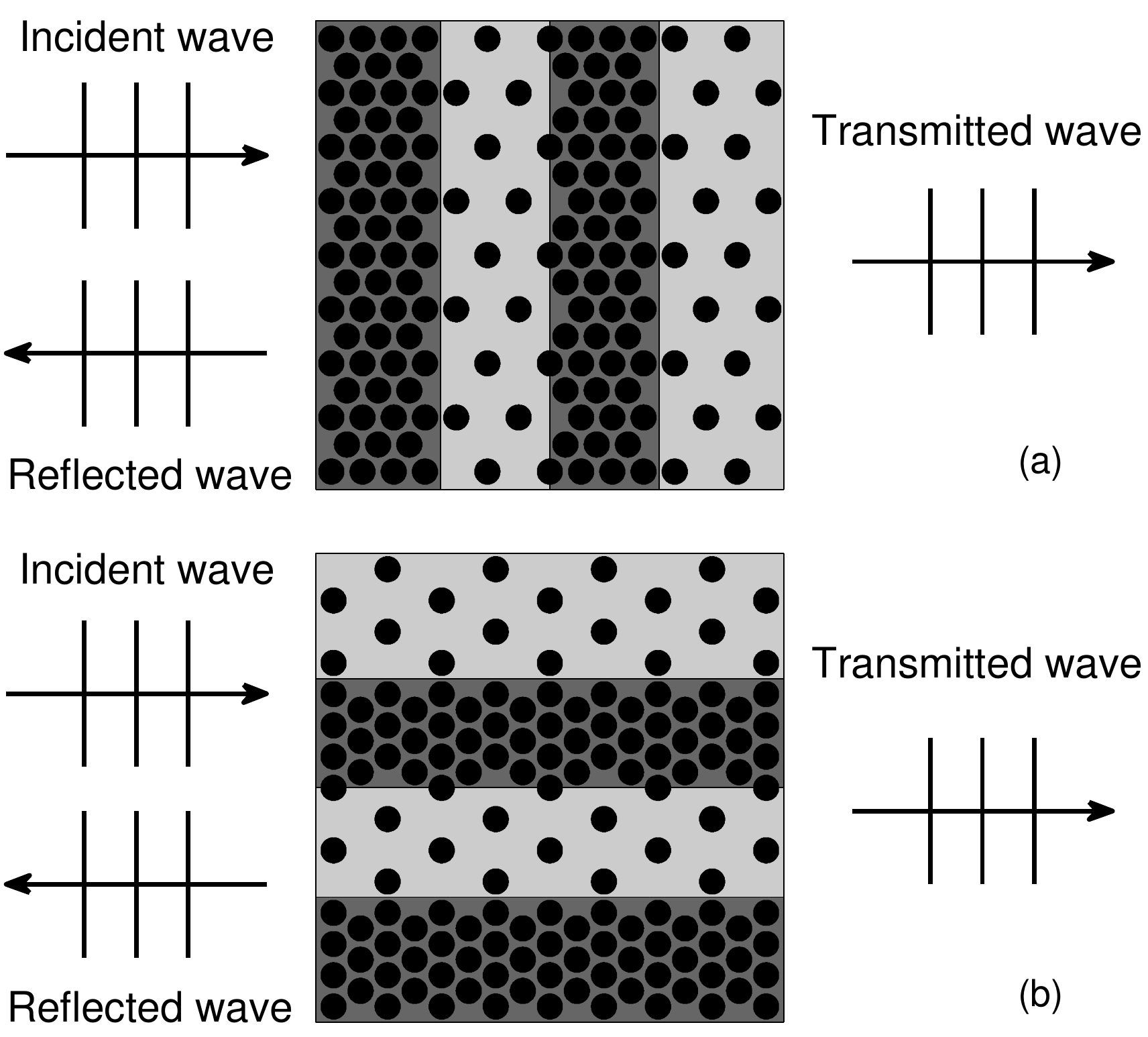}
	\caption{(a) Normal and (b) perpendicular configurations of the anisotropic acoustic metamaterial examined in this work, which consists of two sets of alternating sonic crystal layers.  The two sonic crystal layers behave as effective fluids, which have the homogenized properties of Samples A and C given in Table~\ref{Tab:SampleABC}, and are denoted by dark and light gray, respectively.}
	\label{Fig:AnisoLatticeGeom}
\end{figure}

In this section, the formulation for the effective properties of an acoustic metamaterial with complex anisotropic inertia will be examined, which consists of an alternating-layer arrangement of sonic crystal lattices.  The anisotropy in the inertia arises from differences in the effective density of the homogenized structure at different orientations of the structure.  For the impedance tube testing under investigation in this work, of particular interest is the analysis relating to normal incidence plane waves for two specific configurations, where the impinging wave is either normal or perpendicular to the sonic crystal layers, which are illustrated in FIG.~\ref{Fig:AnisoLatticeGeom}(a) and (b), respectively.

When the acoustic metamaterial is oriented perpendicular to the incident wave, the effective density and bulk modulus of an alternating layer structure is given by the harmonic average of the quantities, namely, \cite{Schoenberg}
\begin{align}
	\rho_{\mathrm{eff}} &= \left[ \frac{d_{1}}{d_{\mathrm{tot}}} \frac{1}{\rho_{1}} +  \frac{d_{2}}{d_{\mathrm{tot}}} \frac{1}{ \rho_{2}} \right]^{-1},  \label{Eq:rho_perp} \displaybreak[0] \\
	\kappa_{\mathrm{eff}} &= \left[ \frac{d_{1}}{d_{\mathrm{tot}}} \frac{1}{\kappa_{1}} +  \frac{d_{2}}{d_{\mathrm{tot}}} \frac{1}{\kappa_{2}} \right]^{-1},  \label{Eq:K_perp}
\end{align}
\noindent where $d_{\mathrm{tot}} \!=\! d_{1} \!+\! d_{2}$, and the subscripts 1 and 2 refer to the first and second alternating fluids layers.  To determine the effective properties of the acoustic metamaterial, a two-step homogenization process will be performed.  First, each sonic crystal lattice will be homogenized to create an effective fluid layer, using the methods described in Section~\ref{Sec:TheoryAnalysisSC2D} for a 2D sonic crystal, or for the results developed for a confined sonic crystal discussed in Section~\ref{Sec:ConfinedTheorySC}.  Second, these effective fluid layers will be homogenized to obtain the effective properties of the acoustic metamaterial in both the normal and perpendicular orientations of the sonic crystal layers.

For a multilayered arrangement of an arbitrary number of fluid layers, this analysis can be performed using the impedance and pressure translation theorems to obtain the input specific acoustic impedance, $Z_{\mathrm{in}}$, and normalized acoustic pressure, $P$, which are given by \cite{Arnott1991,Pierce}
\begin{align}
	Z_{\mathrm{in}}\!(x_{i}) &= Z_{i} \frac{ Z_{\mathrm{in}}\!(x_{i+\!1}) \cos k_{i} d_{i} + j Z_{i} \sin k_{i} d_{i} }{  Z_{i} \cos k_{i} d_{i} + j Z_{\mathrm{in}} \!(x_{i+\!1}) \sin k_{i} d_{i} },  \label{Eq:ImpTransThm} \displaybreak[0] \\
	P(x_{i+\!1}) &= P(x_{i}) \left[ \cos k_{i} d_{i} + j \frac{Z_{i}}{Z_{\mathrm{in}} \!(x_{i+\!1})} \sin k_{i} d_{i} \right]^{-1},  \label{Eq:PressureTransThm}
\end{align}
\noindent where $x_{i}$ is the position of the $i^{\mathrm{th}}$ fluid interface, $d_{i} \!=\! x_{i+\!1} \!-\! x_{i}$ is the thickness of the $i^{\mathrm{th}}$ layer, $k_{i}$ is the wavenumber of the $i^{\mathrm{th}}$ layer, and $Z_{i}$ is the specific acoustic impedance of the $i^{\mathrm{th}}$ layer.  Implementation of Equations~(\ref{Eq:ImpTransThm}) and (\ref{Eq:PressureTransThm}) can be achieved by solving for the input impedance first, and then evaluating the acoustic pressure. Starting from the last layer (which radiates into air) and working backwards yields the input impedance at each successive layer, until the input impedance at the first layer, $Z_{\mathrm{in}}\!(0)$, is determined.  Likewise, the normalized acoustic pressure can then be determined, starting at the first layer and working forward, until the pressure at the last layer is determined, denoted by $P(L)$, where $L$ is the total length of the multilayer structure.  From these two values, the pressure reflection coefficient, $\mathcal{R}$, and transmission coefficient, $\mathcal{T}$, can be determined,
\begin{equation} \label{Eq:TransThmRT}
	\mathcal{R} = \frac{ Z_{\mathrm{in}}\!(0) - Z_{0}}{ Z_{\mathrm{in}}\!(0) + Z_{0}},  \quad \mathcal{T} = P(L).
\end{equation}

The effective homogenized properties of the ensemble structure can be determined using $\mathcal{R}$ and $\mathcal{T}$ for a single effective fluid layer with specific acoustic impedance $Z_{\mathrm{eff}}$, wavenumber $k_{\mathrm{eff}}$ and length $L$.  Using well-known physical acoustic solutions for a single fluid layer \cite{Blackstock}, one can obtain expressions for the effective properties $Z_{\mathrm{eff}}$ and $k_{\mathrm{eff}}$, such that
\begin{align}
	Z_{\mathrm{eff}} &= \frac{1 + \mathcal{R} + \mathcal{T} \! \cos k_{\mathrm{eff}} L }{ j \, \mathcal{T} \! \sin k_{\mathrm{eff}} L },  \label{Eq:Zeff} \displaybreak[0] \\
	k_{\mathrm{eff}} &= \frac{1}{L} \cos^{-1} \! \left[ \frac{ 1 + \mathcal{T}^{2} \!- \mathcal{R}^{2} }{ 2 \mathcal{T} }  \right],  \label{Eq:kLeff}
\end{align}
\noindent where $\cos^{-1}$ denotes the inverse of the cosine function.  A similar result has been previously derived Fokin \emph{et al.} \cite{Fokin2007}, though in this previous work some uncertainty arises due to the periodic but nonunique solution that results from evaluating the $\cos^{-1}$ function in Equation~(\ref{Eq:kLeff}).  Alternatively, Equation~(\ref{Eq:kLeff}) can be evaluated by unwrapping the solution for the $\cos^{-1}$ function, such as by using the method proposed by Baccigalupi \cite{Baccigalupi1999}.

The effective density, sound speed, and bulk modulus can be determined from Equations~(\ref{Eq:Zeff}) and (\ref{Eq:kLeff}), such that
\begin{equation} \label{Eq:K_eff}
	\rho_{\mathrm{eff}} = Z_{\mathrm{eff}} \frac{ k_{\mathrm{eff}} }{\omega},  \quad c_{\mathrm{eff}} =  \frac{\omega}{ k_{\mathrm{eff}} },  \quad \kappa_{\mathrm{eff}} = Z_{\mathrm{eff}} \frac{ \omega }{ k_{\mathrm{eff}}}.
\end{equation}
In the low frequency limit, the expressions for the effective density and bulk modulus for alternating fluid layers oriented in the normal direction reduces to
\begin{align}
	\rho_{\mathrm{eff}} &= \frac{d_{1}}{d_{\mathrm{tot}}} \rho_{1} +  \frac{d_{2}}{d_{\mathrm{tot}}} \rho_{2},  \label{Eq:rho_norm} \displaybreak[0] \\
	\kappa_{\mathrm{eff}} &= \left[ \frac{d_{1}}{d_{\mathrm{tot}}} \frac{1}{ \kappa_{1}} +  \frac{d_{2}}{d_{\mathrm{tot}}} \frac{1}{\kappa_{2}} \right]^{-1},  \label{Eq:K_norm}
\end{align}
\noindent which corresponds to the previously established results by Schoenberg and Sen\cite{Schoenberg} extensively used in anisotropic metamaterial analysis\cite{Torrent2008a}.  Note that in this case the effective bulk modulus reduces to the same value as in the perpendicular orientation given by Equation~(\ref{Eq:K_perp}), so that the anisotropy occurs only in the density in the quasi-static limit.

\subsection{Comparison of results with Comsol} \label{Sec:AnisoInertiaComsol}

\begin{figure}[t!]
	\includegraphics[width=0.99\columnwidth, height=0.7\textheight, keepaspectratio]{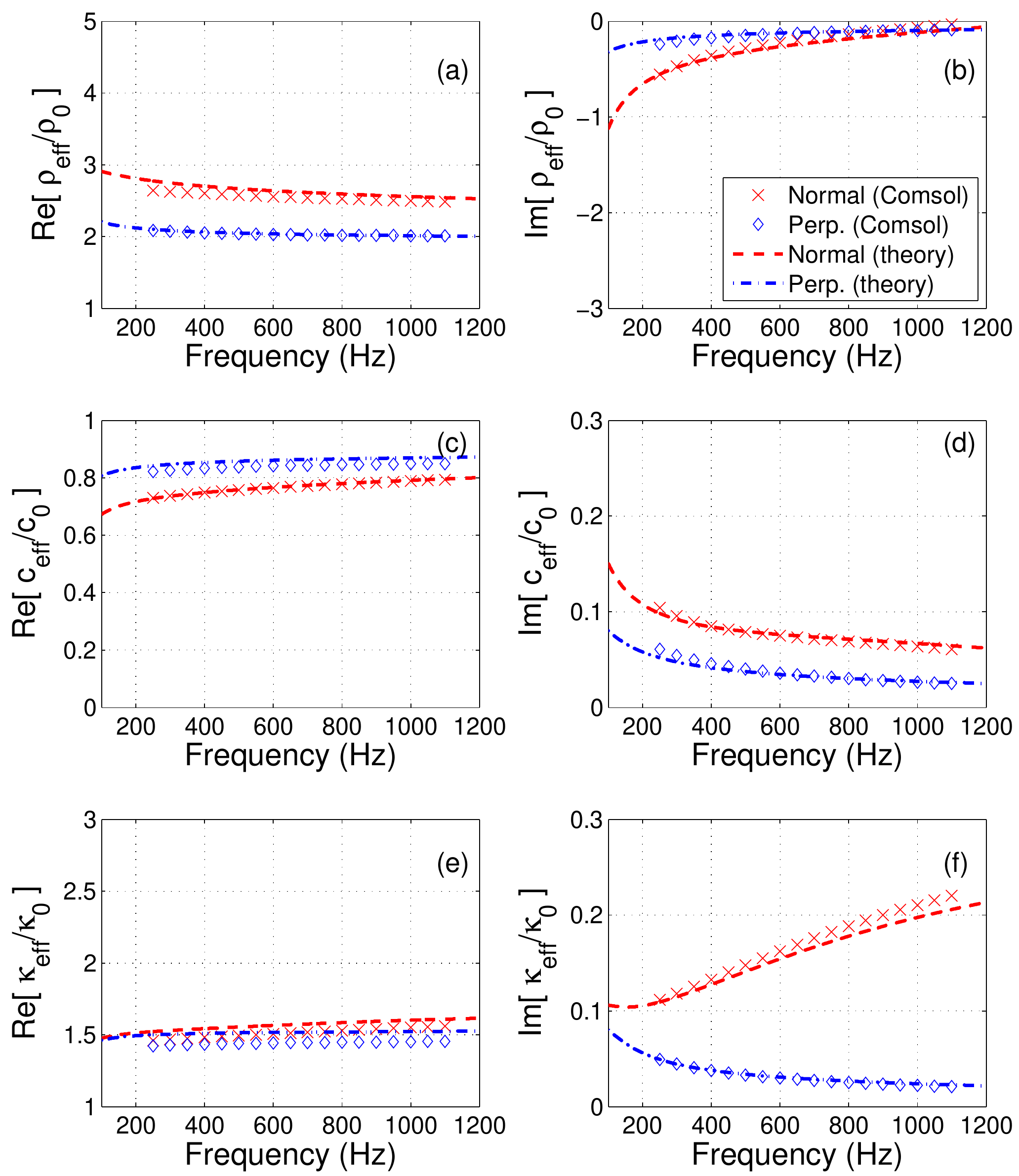}
	\caption{(Color online) Comparison of theoretical results and Comsol for the real and imaginary parts of the density and bulk modulus of a sonic crystal with $r_{0} \!=\! 1$ mm, for the anisotropic acoustic metamaterial configuration illustrated in FIG.~\ref{Fig:AnisoLatticeGeom}.}
	\label{Fig:AnisoInertia2D}
\end{figure}

To further examine the theoretical formulation presented in Section~\ref{Sec:AnisoInertiaTheory}, Comsol was used to determine the effective properties of an acoustic metamaterial with complex anisotropic inertia, for the configuration shown in FIG.~\ref{Fig:AnisoLatticeGeom}.  The results obtained using Comsol are presented in FIG.~\ref{Fig:AnisoInertia2D} for the incident wave normal (x's) and perpendicular (circles) to the sonic crystal layers, consisting of rigid cylinders in air with thermoviscous losses.  Theoretical values for the normal and perpendicular configurations represented by the dashed and dash-dotted lines, respectively, are in excellent agree those obtained with Comsol, for both the real and imaginary part of the effective properties.

In FIG.~\ref{Fig:AnisoInertia2D}(a) and (b), the anisotropy in the density is clearly seen, with the results for the normal direction noticeably higher than that for the perpendicular case.  This anisotropy in the complex density is also apparent in the results for the sound speed illustrated in FIG.~\ref{Fig:AnisoInertia2D}(c) and (d), which shows a similar trend.  In FIG.~\ref{Fig:AnisoInertia2D}(e), there is only a slight difference in the real part of the effective bulk modulus between the two configurations, as expected by the quasi-static results given by Equations~(\ref{Eq:K_perp}) and (\ref{Eq:K_norm}).  Interestingly, a more noticeable difference between the normal and perpendicular configurations is observed in the imaginary part of the bulk modulus shown in  FIG.~\ref{Fig:AnisoInertia2D}(f), a trend that is captured using the theory by retaining the full expressions presented in Section~\ref{Sec:AnisoInertiaTheory}, rather than the quasi-static approximations.  From this observation, it is clear that to correctly account for the losses, it is important to retain the slightly more complicated general expression given by Equations~(\ref{Eq:Zeff})--(\ref{Eq:K_eff}).

\section{Modified formulation for confined sonic crystals} \label{Sec:ConfinedTheorySC}

\begin{figure}[t!]
	\includegraphics[width=0.99\columnwidth, height=0.4\textheight, keepaspectratio]{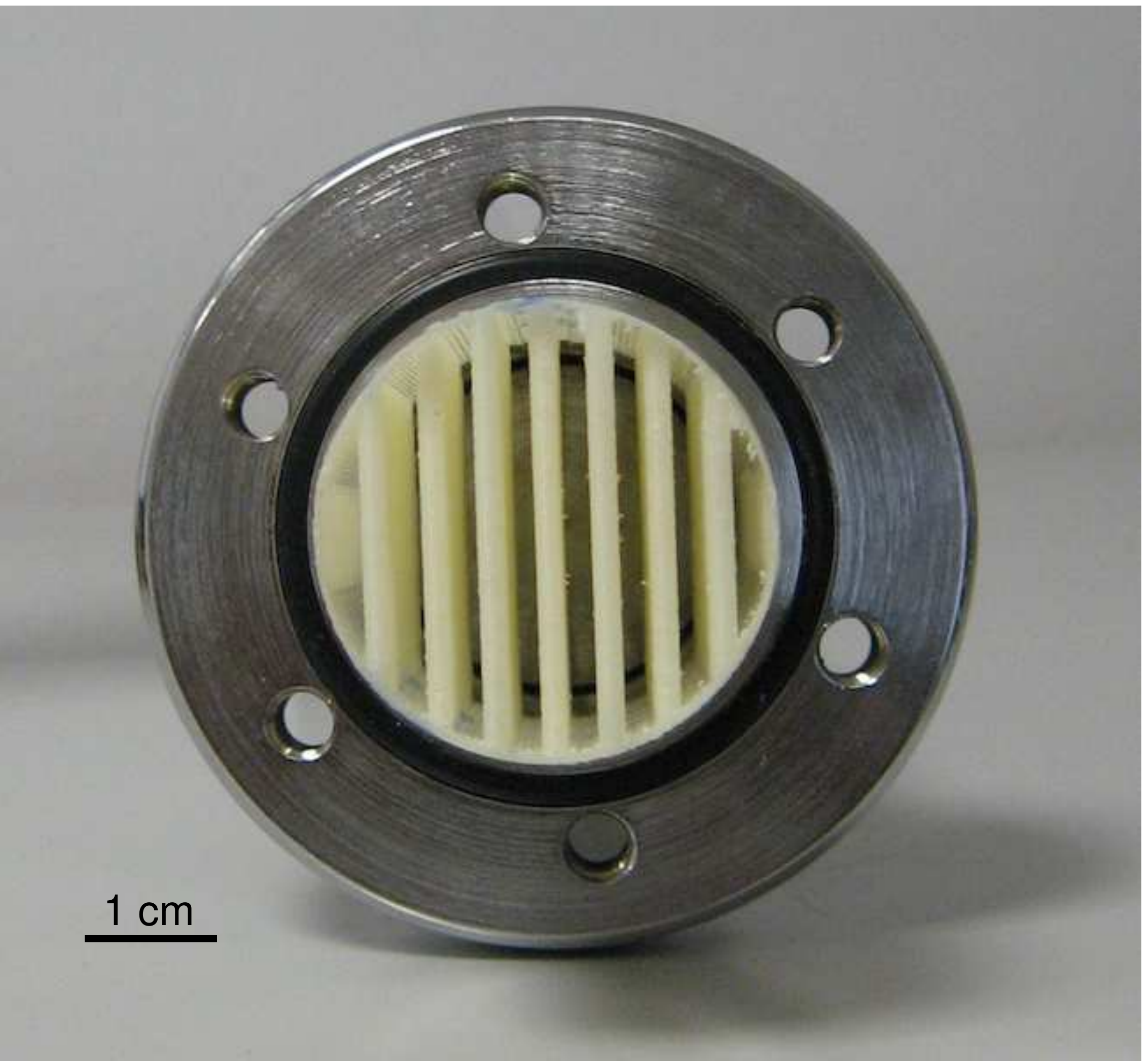}
	\caption{(Color online) Photograph of a confined sonic crystal sample inside an acoustic impedance tube.}
	\label{Fig:ImpTubeSC}
\end{figure}

Through the process of producing a finite sized sonic crystal sample which can be investigated and characterized, a confined sonic crystal or acoustic metamaterial will inevitably be created due to the walls and structure enclosing it. Although creating confined sonic crystals can be done as a design choice, in many practical cases this occurs as a result of using standard acoustic testing techniques, such as an impedance tube.  An illustration of a confined sonic crystal sample situated inside an impedance tube is shown in FIG.~\ref{Fig:ImpTubeSC}.  In this section, modifications to the theoretical results presented in Section~\ref{Sec:TheoryAnalysisSC2D} will be discussed, which can account for these effects.

In general, these modifications from the idealized 2D theory will arise from either the thermoviscous effects within the confined sonic crystal sample itself, or those within the air-filled portion of the impedance tube.  The effects of the impedance tube on the host fluid have been thoroughly examined \cite{MorseIngard,Kinsler}, and for thin boundary layers can be expressed as \cite{Kinsler}
\begin{align}
	\rho_{\mathrm{tube}} &= \rho_{0} \left[ 1 + \frac{1}{2}\left( \frac{\delta}{R_{\mathrm{tube}}} \!+\! \frac{\delta'}{R_{\mathrm{tube}}} \right) \right], \label{Eq:RhoImpTube} \displaybreak[0] \\
	c_{\mathrm{tube}} &= c_{0} \left[ 1 - \frac{1}{2}\left( \frac{\delta}{R_{\mathrm{tube}}} \!+\! \frac{\delta'}{R_{\mathrm{tube}}} \right) \right], \label{Eq:SoundSpeedImpTube}
\end{align}
\noindent where $\delta$ and $\delta'$ are the viscous and thermal boundary layer thicknesses, respectively, and $R_{\mathrm{tube}}$ is the radius of the tube.  Such effects, however, are typically negligible, as the relative contribution from the boundary layer can be mitigated by selecting an appropriately large tube radius for the frequencies under investigation.

The presence of the walls within the sonic crystal sample, as for the case of the air-filled portion of the impedance tube, will also be affected by the viscous and thermal boundary layers emanating from these surfaces.  However, these boundary layers within the sample will be affected by the effective viscous and thermal properties of the sonic crystal sample, which consists of both solid cylinders and the surrounding air.  At moderate to high filling fractions, these effective thermoviscous properties can vary significantly from those of ambient air, leading to observable differences in the effective sonic crystal properties.  In Section~\ref{Sec:ConfinedThermalProp}, expressions for effective thermal and viscous properties are presented, based on effective medium theory.  In Section~\ref{Sec:ConfinedBulkMod}, these effective medium properties are used to determine the appropriate thermal boundary conditions and a revised expression for the dynamic compressibility and bulk modulus for a confined sonic crystal is developed.  Lastly, in Section~\ref{Sec:ConfinedSheath}, some practical obstacles which arise from the fabrication of samples are examined and accounted for in the model, including the effects of a thin plastic sheath around the sonic crystal used for structural support.

\subsection{Effective thermal and viscous properties} \label{Sec:ConfinedThermalProp}
For moderate to high concentrations of inclusions, the bulk properties of the effective medium can be significantly influenced by the number and proximity of the inclusions to one another.  While this has not traditionally been a factor for previous works on unstructured fibrous porous media in air, this effect has been studied quite extensively for elastic composite structures and for fluid emulsions.

To determine the necessary effective properties, recall that the Prandtl number $\mathrm{Pr}$ was used to quantify the relative strength of the thermal and viscous effects, which is determined in Equation~(\ref{Eq:PrandtlDef}) by the specific heat capacity, thermal conductivity and viscosity of the medium.  The effective value for the heat capacity, which is given by the product $\rho \mathrm{C}_{\mathrm{p}}$, is simply \cite{Evans2002}
\begin{equation} \label{Eq:EffHeatCapacity}
	\left[ \rho \mathrm{C}_{\mathrm{p}} \right]_{\mathrm{eff}} = f \left[ \rho \mathrm{C}_{\mathrm{p}} \right]_{\mathrm{inc}} \!\! + (1 \!-\! f) \left[ \rho \mathrm{C}_{\mathrm{p}} \right]_{\mathrm{fluid}},
\end{equation}
\noindent where the subscripts ``inc" and ``fluid" refer to the inclusion and fluid components, respectively.  Material properties for air and the ABS plastic used for the fabricated samples examined in Section~\ref{Sec:Experiment} are given in Table~\ref{Tab:MatProps}.  Note that the density of the cylinder is several orders of magnitude larger than that of air, while the specific heat capacity for most solids are of the same order of magnitude to those for air, so the effective specific heat capacity is $\mathrm{C}_{\mathrm{p,eff}} \! \approx \! \mathrm{C}_{\mathrm{p,inc}}$.  In a similar manner, the thermal conductivity along the axial direction of the effective medium (the vertical direction as shown in FIG.~\ref{Fig:ImpTubeSC}) can be determined using the rule of mixtures, which yields \cite{Christensen}
\begin{equation} \label{Eq:EffTau}
	\tau_{\mathrm{eff}} = f \tau_{\mathrm{inc}} \! + \! (1 \!-\! f) \tau_{\mathrm{fluid}}.
\end{equation}

\begin{table}[t!]
	\begin{ruledtabular}
		\begin{tabular}{lcc}
			\bf{Material properties} & \bf{Air} & \bf{ABS plastic} \\ \hline
			Density, $\rho$ & $1.21 \mathrm{kg}/\mathrm{m}^{3}$ & $1050 \mathrm{kg}/\mathrm{m}^{3}$ \\
			Bulk modulus, $\kappa$ & $142$ kPa & $2.4$ GPa \\
			Shear modulus, $\mu$ & -- & $0.81$ GPa \\
			Compressional wave speed, $c$ & $343$ m/s & $1834$ m/s \\
			Thermal conductivity, $\tau$ & $0.263$ W/m/K & $0.17$ W/m/K \\
			Specific heat capacity, $\mathrm{C}_{\mathrm{p}}$ & $1000$ J/kg/K & $1300$ J/kg/K \\
			Ratio of specific heats, $\gamma$ & $1.4$ & $1.0$ \\
			Viscosity, $\eta$ & $18.5 \, \mu \mathrm{Pa}\!\cdot\! \mathrm{s}$ & -- \\
		\end{tabular}
	\end{ruledtabular}
	\caption{Physical properties of air and ABS plastic used for the fabricated sonic crystal samples examined in Section~\ref{Sec:Experiment}.}
	\label{Tab:MatProps}
\end{table}

In addition to the thermal properties, the effective viscosity will be affected by the presence of the inclusions.  This well-known phenomenon has been traditionally examined for suspensions, and the classic solution for a low concentration of rigid spheres in a viscous fluid is attributed to Einstein, who found that \cite{Einstein}
\begin{equation} \label{Eq:EffViscSpheres}
	\eta_{\mathrm{eff}} = \eta \left(1 + \frac{5}{2} f \right).
\end{equation}
\noindent Although this expression is extensively used, its applicability is limited to objects with a spherical shape.  For the case of rigid cylinders in a viscous fluid, the effective viscosity can be determined by analogy with elastic composites.  Specifically, it has been observed that the results of rigid inclusions in elastic solids share the same fundamental mathematical structure.  Thus, by examining the effective shear modulus for such a composite, an expression for the effective viscosity can be obtained by taking the limiting case of a perfectly incompressible material \cite{Christensen}.  For an elastic solid composite with parallel rigid fibers, the effective shear modulus in the transverse direction, $\mu_{\mathrm{eff}}$, is \cite{Christensen}
\begin{equation} \label{Eq:EffShearMod}
	\mu_{\mathrm{eff}} = \mu \left[1 + \frac{4 f (1 \!-\! \nu)}{3 \!-\! 4 \nu} f \right],
\end{equation}
\noindent where $\mu$ and $\nu$ are the shear modulus and Poisson's ratio of the host elastic material.  In the limit of an incompressible material, $\nu \! \to \! \frac{1}{2}$, and therefore by analogy $\mu \! \to \! \eta$, so Equation~(\ref{Eq:EffShearMod}) reduces to an expression for the effective viscosity of rigid parallel cylinders in a viscous fluid,
\begin{equation} \label{Eq:EffViscCyl}
	\eta_{\mathrm{eff}} = \eta \left(1 + 2 f \right).
\end{equation}
Although this linearly proportional relationship to the filling fraction is similar in form to that derived by Einstein, the increase with the filling fraction is slightly less due to the different geometry.  In both cases, it is observed that the presence of the inclusions lead to an increase in the observable viscosity of the effective medium.  The effective density for a confined sonic crystal is given by the same expression as for the unconfined case described by Equation~(\ref{Eq:Rho_SC}), except with an increase in the effective viscosity, as described by Equation~(\ref{Eq:EffViscCyl}).  Note that this increased viscosity corresponds to higher flow resistivity according to Equation~(\ref{Eq:SigmaDef}), and therefore an increase in the imaginary part of the density and ultimately higher losses.

\subsection{Effective bulk modulus for confined sonic crystals} \label{Sec:ConfinedBulkMod}
For a confined sonic crystal, the presence of the surrounding surfaces will affect how the thermoviscous boundary layers interact with the cylinders and the resulting effective bulk modulus of the homogenized structure.  Although the fundamental equations are the same as for the unconfined sonic crystal, the cylinders within the confined sonic crystal will experience a different thermal boundary condition at $r \!=\! R$, the outer radius of the unit cell.  For high filling fraction applications relative to the boundary layer thickness, the outer unit cell boundary conditions are often set equal to those of the effective medium, to compensate for the net interaction from the surrounding cylinders \cite{Evans2002}.  This approach of treating a unit cell surrounded by an effective homogenized medium is also utilized extensively in effective medium theory for elastic solids \cite{Christensen}.

In the idealized 2D expressions developed in Section~\ref{Sec:TheoryAnalysisSC2D}, it was appropriate to assume that the interactions between cylinders could be neglected, since the thermal and viscous effects were confined to the relatively thin boundary layers close to each cylinder, leading to adiabatic thermal boundary conditions at $r \!=\! R$.  For the confined sonic crystal, however, viscous and thermal boundary layers emanate from the walls of the confining structure, with thermoviscous properties of the effective medium.  Even at low filling fractions, these boundary layers can be significantly larger than those in air for thermally conductive cylinders, and under these circumstances it is appropriate to apply thermal boundary conditions at $r \!=\! R$ equal to those of an effective homogenized sonic crystal medium.  In this case, the boundary condition with thermal conduction across the interface can be described in a similar manner to Equation~(\ref{Eq:ThermCondBC}), 
\begin{equation} \label{Eq:EffMedThermalBC}
	2 \pi R \tau \! \left. \frac{\partial T}{\partial r} \right|_{r \!=\! R} \!\!  = j \omega \, \pi R^{2} \, \rho_{\mathrm{eff,static}} \mathrm{C}_{\mathrm{p,eff}} T(R),
\end{equation}
\noindent where $\tau$ is the thermal conductivity of the fluid (air) and $\rho_{\mathrm{eff,static}}$ is the static density of the effective medium.  For an effective homogenized medium surrounding the unit cell containing even low to moderate filling fractions of thermally conducting cylinders, the right hand side will be significantly larger and will yield an approximately isothermal boundary condition, $T(R) \! \approx \! 0$.   By applying isothermal boundary conditions and repeating the analysis to solve for the temperature increase, one finds the expression for the dynamic compressibility becomes
\begin{equation} \label{Eq:C_HighFF}
	C_{\mathrm{SC,conf}} = 1 - (\gamma-1) \frac{2 f}{(1 \!-\! f)} \bar{H}_{\mathrm{conf}},
\end{equation}
\noindent where
\begin{align}
	\bar{H}_{\mathrm{conf}} =& \bigg\{ \Big[ k_{T}r_{0} J_{1}(k_{T}r_{0}) \!-\! k_{T}R J_{1}(k_{T}R) \Big] \Big[H_{0}^{(2)}(k_{T}r_{0}) \!-\! H_{0}^{(2)}(k_{T}R)\Big] \notag \displaybreak[0] \\
	&  \!-\! \Big[J_{0}(k_{T}r_{0}) \!-\! J_{0}(k_{T}R)\Big] \Big[ k_{T}r_{0} H_{1}^{(2)}(k_{T}r_{0}) \!-\! k_{T}R H_{1}^{(2)}(k_{T}R) \Big] \bigg\} \notag \displaybreak[0] \\
	&\bigg\{ J_{0}(k_{T}r_{0})\Big[ H_{0}^{(2)}(k_{T}r_{0}) \!-\! H_{0}^{(2)}(k_{T}R)\Big]  \notag \displaybreak[0] \\
	& \qquad \qquad  \!-\! \Big[J_{0}(k_{T}r_{0}) \!-\! J_{0}(k_{T}R)\Big] H_{0}^{(2)}(k_{T}r_{0}) \bigg\}^{-1}. \label{Eq:H_bar}
\end{align}
Therefore, the bulk modulus of a confined sonic crystal can be described by Equation~(\ref{Eq:BulkMod_SC}), with the dynamic compressibility given by Equations~(\ref{Eq:C_HighFF}) and (\ref{Eq:H_bar}).  Although Equations~(\ref{Eq:H_bar}) is a somewhat complicated expression, it is expected that a decrease in the effective bulk modulus will be observed due to the change from adiabatic to isothermal conditions.

\subsection{Effects of sample sheath} \label{Sec:ConfinedSheath}
Although the expressions given above fully describe the effective density and bulk modulus of a confined sonic crystal, in this section the effects of the plastic sheath used to provide structural support for the fabrication and experimental testing of the samples will be discussed.  Although the plastic sheath is quite thin compared with the radius of the impedance tube (as seen in FIG.~\ref{Fig:ImpTubeSC}), the presence of this sheath leads to two main effects: an increase in the measured effective acoustic impedance, and it leads to a layer of air which increases the effective density and decreases the effective bulk modulus the sample.

The total specific acoustic impedance due to the sonic crystal with the plastic sheath is given by
\begin{equation} \label{Eq:EffZsheath}
	Z_{\mathrm{eff,sheath}} = Z_{\mathrm{eff}} \left[ 1 - \phi_{\mathrm{sheath}} \right]^{-1} ,
\end{equation}
where $Z_{\mathrm{eff}}$ is the specific acoustic impedance of the homogenized sonic crystal and $\phi_{\mathrm{sheath}} \!=\! l_{\mathrm{sheath}} / R_{\mathrm{tube}}$, with $l_{\mathrm{sheath}}$ and $R_{\mathrm{tube}}$ denoting the sheath thickness and the radius of the tube, respectively.  Thus, $Z_{\mathrm{eff,sheath}}$ represents the impedance measured within the impedance tube, and the resulting effective density and bulk modulus are likewise scaled as
\begin{align}
	\rho_{\mathrm{eff,sheath}} &= \rho_{\mathrm{eff}} \left[ 1 - \phi_{\mathrm{sheath}} \right]^{-1}, \label{Eq:EffRhoSheath} \displaybreak[0] \\
	\kappa_{\mathrm{eff,sheath}} &=  \kappa_{\mathrm{eff}} \left[ 1 - \phi_{\mathrm{sheath}} \right]^{-1}. \label{Eq:EffBulkModSheath} 
\end{align}

In addition to the sheath encasing the sonic crystal, a slight lip (where the sheath extended slightly past the cylinder) was present due to the fabrication process.  The result of this lip is a thin layer of air adjacent to the front and back of sonic crystal sample.  The observed effective properties can be quantified by assuming that the resulting thin air layer acts like an acoustic lumped element, in which case
\begin{align}
	\rho_{\mathrm{eff,lip}} &= \rho_{\mathrm{eff}} + \rho_{0} \phi_{\mathrm{lip}}, \label{Eq:EffRhoSheathLip} \displaybreak[0] \\
	\kappa_{\mathrm{eff,lip}} &= \left[ \frac{1}{\kappa_{\mathrm{eff}}} +  \frac{ \phi_{\mathrm{lip}} }{\kappa_{0}} \right]^{-1}, \label{Eq:EffBulkModSheathLip} 
\end{align}
\noindent where $\phi_{\mathrm{lip}} \!=\! l_{\mathrm{lip}} / R_{\mathrm{tube}}$.

From Equations~(\ref{Eq:EffRhoSheath})--(\ref{Eq:EffBulkModSheathLip}) it can be observed that the sheath and the sheath lip will increase the measured effective density.  For the bulk modulus, the sheath itself will lead to an increase due to the increase in the specific acoustic impedance, though the presence of the sheath lip acts to reduce the effective bulk modulus.  Although which factor dominates depends on the precise thicknesses of the sheath and sheath lip, it is clear from Equation~(\ref{Eq:EffBulkModSheathLip}) that the observable effective bulk modulus resulting from the sheath lip will depend on the relative magnitude of the effective bulk modulus of the sonic crystal compared with that of the ambient air.  Thus, the presence of the sheath lip will be amplified as the filling fraction increase, and therefore one would expect this effect to dominate at higher filling fractions, where $\kappa_{\mathrm{eff}} \! \gg \! \kappa_{0}$.

\section{Experimental results} \label{Sec:Experiment}

To verify the theoretical model for confined sonic crystals, several samples were fabricated and experimentally tested using a standard circular cross-section acoustic impedance tube.  The inner diameter of the impedance tube is 3.5 cm, and the end of the tube is terminated with fiber glass insulation to provide an anechoic termination.  Noise is generated and transmitted using a electromechanical driver, and measured using 0.50 inch (1.27 cm) diameter G.R.A.S. condenser microphones.  The microphones are arranged in a standard 4-microphone configuration\cite{Salissou2010}, allowing for the magnitude and phase of both the reflection and transmission pressure coefficients to be directly determined using a transfer-matrix method \cite{Song2000}, from which the complex impedance and wavenumber were obtained for the range 300--2000 Hz.  These values correspond to the same effective properties described for the theoretical model given by Equations~(\ref{Eq:Zeff}) and (\ref{Eq:kLeff}).
 
\subsection{Isotropic inertia}
The sonic crystal samples were created using a commercial 3D printer out of ABS plastic for several different configurations, which are listed in Table~\ref{Tab:SampleABC} and cover a wide range of filling fractions.  The three sonic crystal samples described in Table~\ref{Tab:SampleABC}, consisting of a single uniform arrangement with constant lattice parameter, were constructed to verify the results of the modeling of the confined sonic crystals, based on the formulations presented in Section~\ref{Sec:ConfinedTheorySC}.  A photograph of a sample mounted in the impedance tube is shown in FIG.~\ref{Fig:ImpTubeSC}.  In this figure, the thin plastic sheath surrounding the sonic crystal   of the test sample can be seen, which was necessary for structural support to ensure the cylinders remained properly aligned.

\begin{figure}[t!]
	\includegraphics[width=0.99\columnwidth, height=0.7\textheight, keepaspectratio]{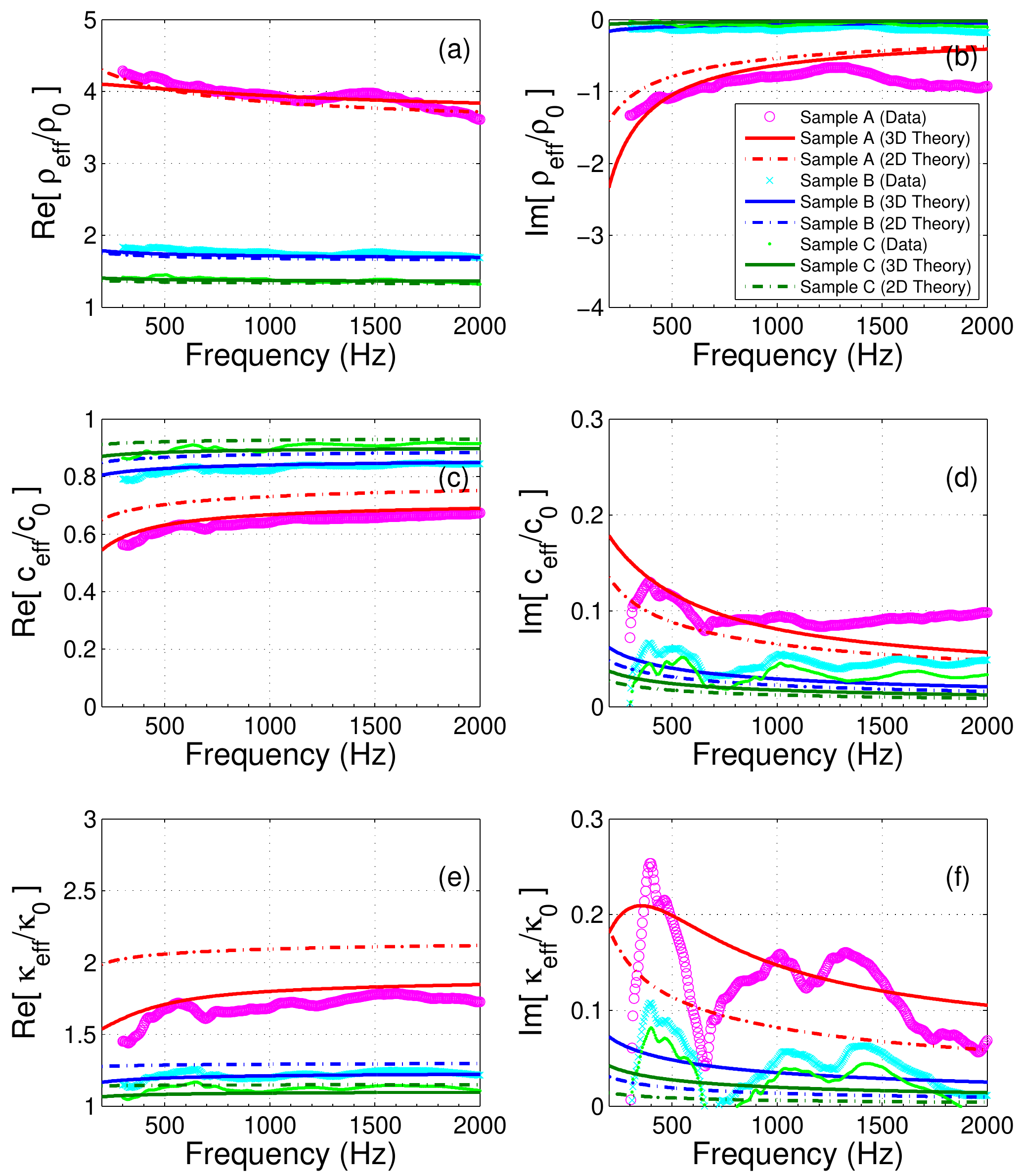}
	\caption{(Color online) Comparison of theoretical results and experimental impedance tube data for the real and imaginary parts of the density and bulk modulus of a sonic crystal with $r_{0} \!=\! 1$ mm, for Sample A, B, and C.}
	\label{Fig:SCdata3D}
\end{figure}

Figure~\ref{Fig:SCdata3D} shows the experimental results for the complex density, sound speed and bulk modulus for Samples A, B, and C.  For comparison, two theoretical models of the sonic crystal samples are presented: the first being the 2D sonic crystal model (also shown in FIG.~\ref{Fig:ModelComp2D}), and second including the modifications for the confined sonic crystal with the effective thermal properties and isothermal boundary conditions for the dynamic compressibility described by Equations~(\ref{Eq:C_HighFF}), (\ref{Eq:EffViscCyl}), and (\ref{Eq:H_bar}).  For the real and imaginary parts of the density shown in FIG.~\ref{Fig:SCdata3D}(a) and (b), there is excellent agree between both models and the experimental results, with only a slight deviation observed with the unconfined 2D sonic crystal model at the highest filling fraction (Sample A).  For the complex sound speed and bulk modulus shown in FIG.~\ref{Fig:SCdata3D}(c)--(f), there is a much more significant difference between the modeled results, resulting from the different thermal boundary conditions used to derive the expressions for the dynamic compressibility, and thus the bulk modulus.  In particular, it is observed that the 2D sonic crystal model, which was in excellent agreement with the 2D results presented in FIG.~\ref{Fig:ModelComp2D}, yields a bulk modulus which has a significantly higher real part, with a correspondingly lower imaginary part, than the experimental data.  This trend is also observed in the sound speed data as well.  However, the theoretical formulation with the modifications for the confined sonic crystal correctly accounts for this decrease in the bulk modulus and increase in the losses (characterized by the imaginary part of the properties), and is in excellent agreement with the experimental results for the entire range of filling fractions examined.

\begin{figure}[t!]
	\includegraphics[width=0.99\columnwidth, height=0.7\textheight, keepaspectratio]{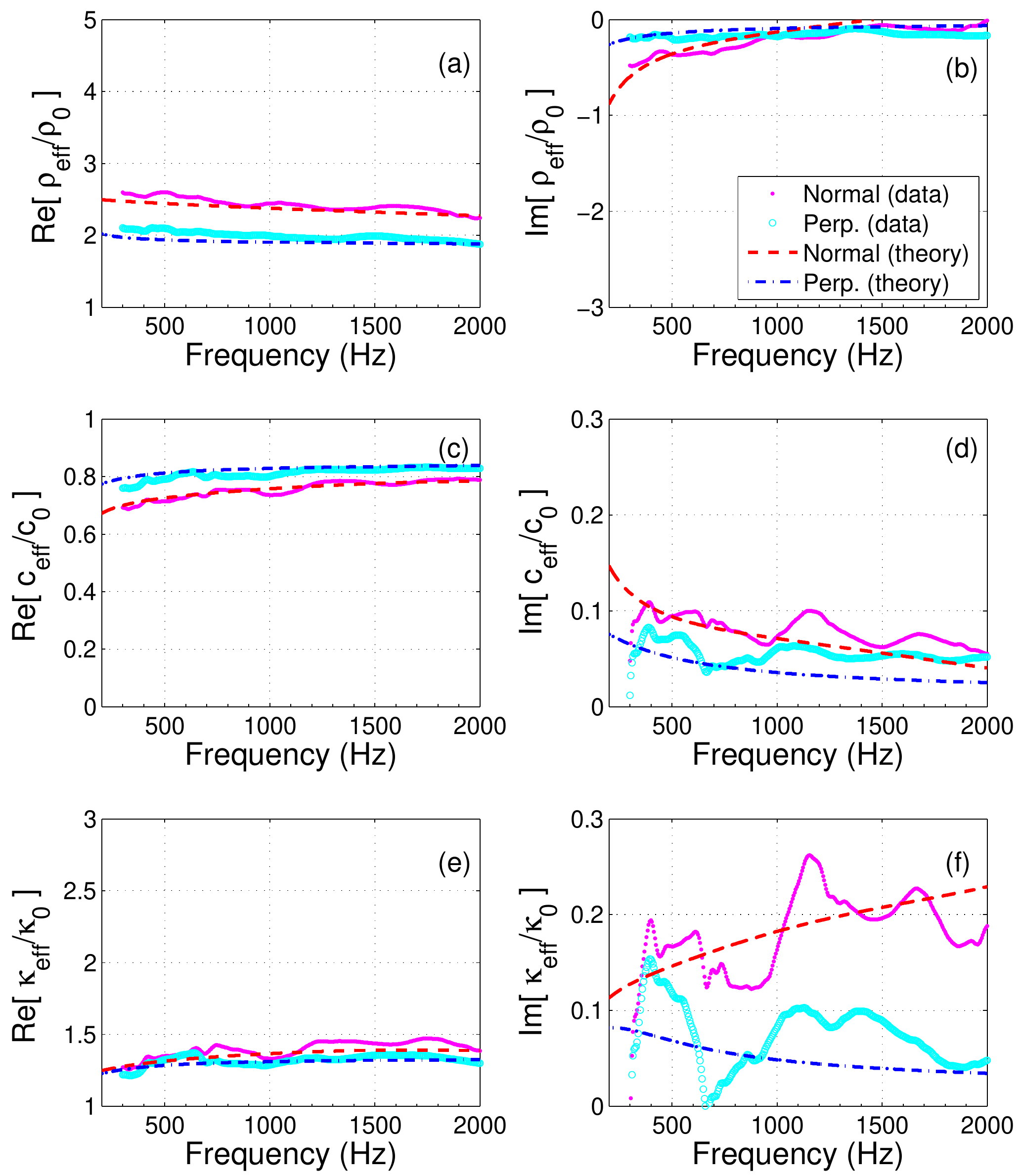}
	\caption{(Color online) Comparison of theoretical results and experimental impedance tube data for the real and imaginary parts of the density and bulk modulus of a sonic crystal with $r_{0} \!=\! 1$ mm, for the anisotropic acoustic metamaterial configuration illustrated in FIG.~\ref{Fig:AnisoLatticeGeom}.}
	\label{Fig:AnisoInertiaData3D}
\end{figure}

\subsection{Anisotropic inertia}
For the realization of the acoustic metamaterial with complex anisotropic inertia, two samples were constructed using the arrangement illustrated in FIG.~\ref{Fig:AnisoLatticeGeom}(a) and (b) to demonstrate the anisotropy in the normal and perpendicular directions, respectively.  As seen in FIG.~\ref{Fig:AnisoLatticeGeom}, these samples consist of two sets of alternating sonic crystal layers, containing a high filling fraction layer (dark) and low filling fraction (light).  The same experimental setup was used to test these acoustic metamaterial samples, the results for which are presented in FIG.~\ref{Fig:AnisoInertiaData3D}.  For comparison, theoretical results using Equations~(\ref{Eq:Zeff}) and (\ref{Eq:kLeff}) to calculate the effective properties of the structure, with the homogenized layer properties based on those for confined sonic crystals (corresponding to the theoretical results presented for Sample A and C in FIG.~\ref{Fig:SCdata3D}).  Upon examination of the data shown in FIG.~\ref{Fig:AnisoInertiaData3D}, excellent agreement is observed between the theoretical and experimental results.  This corresponds to an accurate description of both the real and imaginary parts of each effective property, for each orientation.  The anisotropic inertia for this acoustic metamaterial is observed in the data through the significantly different values in the real and imaginary parts of the effective density for the normal and perpendicular orientations.  These trends are precisely captured by the theoretical results, for the complex values of the density, sound speed and bulk modulus.

\section{Conclusions} \label{Sec:Conclusion}

In this work, we present theoretical and experimental results for the consideration of thermal and viscous losses on the performance of sonic crystals, with filling fractions much larger than traditional porous absorbers.  Due to the ordered microstructure, expressions for the complex effective parameters of sonic crystals can be written with no unknown or empirical coefficients.  In addition, it is shown that they can be completely characterized by only the filling fraction and normalized boundary layer thickness.  From these results, parametric plots are developed and examined, and highlight desirable characteristics for the enhancement of sound absorption, including loss factors near unity and low bulk moduli.  The effects of a confining structure around a sonic crystal lattice is examined theoretically and experimentally, with the results showing excellent agreement over a wide range of filling fractions and frequencies.  A formulation for acoustic metamaterials with complex-valued effective material properties is presented, making use of these confined sonic crystal properties, which is also in excellent agreement with the theoretical model.  Although only a relatively simple configuration  was examined, the anisotropic acoustic metamaterial and confined sonic crystal formulations discussed and developed here have the potential for more complicated designs, enabling the construction of effective fluid sound absorbers that have anisotropy in both the material properties and absorption characteristics, as well as the potential for creating soft acoustic metamaterials with enhanced sound absorption performance.

\section*{Acknowledgements}
This work was supported by the U.S. Office of Naval Research (Award N000141210216) and by the Spanish \emph{Ministerio de Economia y Competitividad} (MINECO) under contract No. TEC2010-19751.


%

\end{document}